\documentclass[12pt]{article}
\usepackage{graphicx}
\usepackage{url}
\begin{document}

\def\b{\bigskip}
\def\s{\smallskip} \def\c{\centerline}  \baselineskip=15 pt  \c {\bf Shannon entropy and blackbody radiation }\b \c {Italo Vecchi} \c {Via Leoncorno 5 - 44100 Ferrara -
Italia} \c { email: vecchi@isthar.com } \b 

{ \bf Abstract}: {\sl A response to criticism of  the author's previous paper \url{http://xxx.lanl.gov/abs/quant-ph/0002084}.}\b

 This note elaborates on  the author's  previous paper [1], where Planck's radiation law was examined as an instance of a macroscopic quantum phenomenon.  The goal of the following remarks is just to stress the absolute  simplicity of the  argument in [1].\s 

 The main point in [1] may be summed up  recalling that Planck's radiation law is  determined by maximising the Gibbs entropy on the discrete spectrum of the energy operator.  In the case of Planck's radiation law  therefore  the structure of the macroscopic phenomenon is associated to a "preferred basis",  the energy eigenvectors, which corresponds to the  energy  operator. Planck's radiation law applies also to "closed" systems, so that the choice of the operator depends on the observer and cannot be attributed to any interaction with the environment. It was stressed in [1] that measurement operators with different spectra induce different distribution laws. The Jeans-Raleigh distribution law, for example,  is obtained
maximizing the Gibbs entropy on continuos  spectra.  
Other spectra, i.e. other operators, which  can be easily manufactured , yield different distributions. The macroscopic properties of the radiation field depend therefore on the operator´s spectrum, i.e. on the measurement outcomes of an observer measuring the system with the given operator.  \s
 
The above reduces to the rather obvious statement that  the density of a Gibbs canonical ensemble (e.g. Planck's radiation law) is  relative to an observable, i.e. to the spectrum of a measurement operator. Different measurement operators yield different densities.\s

 It is worth recalling that the  density maximises entropy on observables with a given average. In other words, given a set of measurement outcomes with a given average, the density corresponds to the distribution of measurement outcomes which maximises entropy. In the case of Planck´s radiation law ,  since  energy is the measurement operator that is being
considered, the average corresponds to temperature. Gibbs densities are calculated  for observables with a given average. The  density applies to the observable, i.e. to the sets of measurement outcomes with the given average.   \s

If we consider the density matrix corresponding to the system, it should be clear tha the entropy to be maximised is not the von Neumann entropy $ Tr{\rho log(\rho)} $ but  the Shannon entropy  obtained from the observer's measurement  probability distribution relative to the operator´s spectrum. The entropy that is being maximised is not basis independent, sice the it applies to a given spectrum.   \s

We can clarify this point  considering again  Planck's law. The key point here is that discrete energy is the measurement operator that is being considered. The requirement  that the mean energy, i.e. temperature,  is fixed  amounts to imposing  a condition on the averaged  measurement outcomes of the observer's operator, i.e. on the discrete energy operator itself. The distribution applies to ensembles which satisfy the constraint.   The radiation field's diagonal density matrix is the result of maximising  entropy  under a constraint which, as pointed out above, applies to  a given operator associated to the observer. The entropy being maximised is relative to this operator and is therefore a Shannon entropy, obtained under  constraints  on the operator's measurement outcomes ( i.e. the mean energy is fixed). The resulting density matrix describes a mixture, as Planck distribution does not determine  a state of the system, but a  macroscopic property of an ensemble relative to an observer. \s

Summing up the above one can say that  the density of a Gibbs canonical ensemble for an observable maximises the Shannon entropy for that observable. Planck' radiation law  maximises the Shannon entropy relative to the  energy operator. The Jeans-Raleigh law maximises the Shannon entropy relative to an   operator with a continuous spectrum. \s

The same argument can be appiled to the diagonalisation of the density matrix which is  the hallmark of decoherence theory. In [1]  the author points out that the choice of the basis in which the diagonalisation takes place does not depend on the environment but on the observer,  since physical systems do not chose a basis. Indeed the diagonalisation of the density matrix corresponds to the maximisation of the Shannon entropy for the operator  whose measurement outcomes are constrained.
For Planck's radiation law the constraint is that the mean energy is fixed. For Schroedinger's Cat the constraint is that the cat will be there when the box is opened, dead or alive. This constraint applies to the eigenvalues of the density matrix in the given base. If the matrix is diagonal, it's a condition on the matrix's  trace. \s 

The above casts an interesting light on decoherence.  Quite simply, diagonalisation of the density matrix corresponds to the loss of information under the constraint that some measurement outcomes are preserved. For example localisation, i.e. the vanishing of spatial superpositions, turns out to encode only the fact that the observer knows that his steel ball has ended up somewhere. The resulting diagonal density matrix maximises entropy on the measurement  outcomes of the position operator under this constraint.  \s

  I would like to add some general remarks, which, although not entirely original ( they are more or less implicit in the von Neuman-Stapp approach), may be helpful. The issue of decoherence and in general all the issues related to quantum measurements appear riddled with problems whose nature is semantic. In this context the problem of the meaning of words acquires direct physical relevance. An instance of this problem is the word "macroscopic"
which is widely used but whose meaning appears often ambiguos.  I propose to define "macroscopic" as "relative to a measurement outcome", since all "macroscopic"  phenomena are perceived by the observer as the result of an act of measurement. Moreover since any measurement depends on an operator, which is associated to an abserver,  it follows that "macroscopic" means also "relative to the measurement outcomes of a given operator associated to an observer". 
The "macroscopic" properties of physical systems therefore depend on the operators associated to the observer. A simple instance of this is the black body radiation, as shown above. We observe  blackbody radiation in accordance with Planck's law because we use the discrete energy operator to observe it. 

\b

 \c{ \bf References} \b

[1] I.Vecchi, {\sl Decoherence and Planck`s Radiation Law} at \url{http://xxx.lanl.gov/abs/quant-ph/0002084}. \s

\end{document}